\documentclass
[preprint,showpacs,preprintnumbers,amsmath,amssymb,apsrev]{revtex4}%
\usepackage{graphicx}
\usepackage{dcolumn}
\usepackage{bm}
\usepackage{amsmath}
\usepackage{amsfonts}
\usepackage{amssymb}%
\begin{document}
\preprint{preprint}
\title{Controllable Flux Coupling for the Integration of Flux Qubits}
\author{C. Cosmelli}
\affiliation{INFN and Dipartimento di Fisica, Universit\`a di Roma La Sapienza, P.le A. Moro 5,
00185 Roma, Italy}
\author{M. G. Castellano, F. Chiarello, R. Leoni, D. Simeone, G. Torrioli}
\affiliation{ Istituto di Fotonica e Nanotecnologia, CNR, Via Cineto Romano 42, 00156 Roma, Italy}
\author{P. Carelli}
\affiliation{Dipartimento di Ingegneria Elettrica, Universit\`a dell'Aquila, Monteluco di
Roio, 67040 L'Aquila, Italy}
\date{\today}

\begin{abstract}
We show a novel  method for  controlling the coupling of flux-based qubits by means of a
 superconducting transformer with variable flux transfer function. 
The device is realized by inserting a small hysteretic dc
SQUID with unshunted junctions, working as a Josephson junction with flux-controlled critical
current, in parallel to a superconducting transformer; by varying the magnetic flux coupled to the
dc-SQUID, the transfer function for the flux coupled to the transformer can be varied. Measurements
carried out on a prototype at $4.2$ K show a reduction factor of about $30$ between the ``on'' and the ``off'' states.
We discuss the system characteristics and the experimental results.

\end{abstract}

\maketitle



Recently, different types of qubits, all based on Josephson junctions, have been experimentally
demonstrated. Flux~\cite{friedman2000,chiorescu2003}, phase~\cite{martinis2002,yu2002} and
phase-charge~\cite{vion2002} qubits have been operated as single devices, while charge state qubits
have also been used in an entangled couple, showing quantum-coherent behavior~\cite{pashkin2003}
and operation as a conditional gate~\cite{yamamoto2003}. In the implementation of a system of
entangled qubits, one of the challenges is the realization of a connection between different qubits
that fulfills the various constraints imposed by the correct qubit operation. The connection should
be non dissipative, otherwise the fluctuations related to its dissipation will destroy the coherent
state of the connected qubits; this forbids the use of resistive elements or elements that are
dissipative even for a short period of time. It should allow a fast switching, namely its switching
time should be much faster than the time related to the clock period. It should be simple and
reliable, to be integrable with a large numbers of qubits, and the related implementation should be
a well-established technology with a very high degree of reliability. Besides, the coupling
strength of the connection should be varied from  outside, allowing to turn the coupling on
and off whenever needed; it must be noted that a scheme with untunable couplings has been
proposed~\cite{zhou2002} but not yet implemented in practical realizations.

In order to couple flux qubits, it comes natural to use superconducting
transformers. Two schemes for achieving coupling control have been presented
recently. In the INSQUID~\cite{insquid}, which was originally ideated for
readout, the flux qubit is placed inside the dc-SQUID of a double-SQUID. The
tunable transformer of ref.~\cite{Filippov2003}, instead, is conceived for
gradiometric flux qubits like that of ref.~\cite{friedman2000} and is based on the
balancing of a gradiometric transformer by means of two small dc-SQUIDs
inserted in the transformer branches.

In this letter we propose a Controllable Flux Coupling (CFC) system, suitable for the connection 
of one or more flux qubits. The CFC basic idea  is to use a superconducting flux transformer, 
modified with the  insertion of a small hysteretic dc SQUID that
behaves as a Josephson junction with tunable critical current. By modulating
the SQUID critical current by means of an external magnetic field, it is
possible to control the flux transfer function  through the transformer and therefore the coupling. Compared to other proposed schemes,
the CFC  has the advantage of being easily coupled to a flux
qubit through inductively coupled coils, without requiring a particular geometry for the qubit to be read out.  

\begin{figure}
\includegraphics{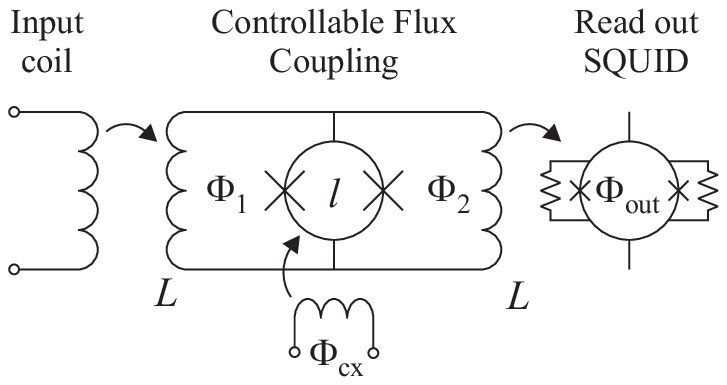}\caption{Schematic of the CFC circuit. The main part is the central gradiometric transformer, with arms of inductance $L$, modified by an inner dc-SQUID of smaller inductance $l$; the coordinates are the magnetic fluxes $\Phi_1$ and $\Phi_2$. A signal flux is applied on the left side and the CFC response is read out by the SQUID on the right side. The control on the transmitted flux is achieved by coupling a flux $\Phi_{cx}$ to the inner dc-SQUID.}
\label{fig:scheme}
\end{figure}

Our variable transformer (Fig.~\ref{fig:scheme}) consists of two arms of inductance $L$, in
parallel with a dc-SQUID that performs the control of the magnetic flux transfer. The inner
dc-SQUID is a loop of inductance $l\ll L$, interrupted by two Josephson junctions of  critical current $I_{0}$ and capacitance $C$; its dynamics is described by the differences of the superconducting phase across the Josephson junctions, which
are linked by the fluxoid equation to the magnetic fluxes $\Phi_{1}$ and $\Phi_{2}$ in the left and
in the right loops. 

Input flux is coupled to the left side through an inductively coupled coil, with flux transforming ratio ${\cal R}_{1}$; the flux appearing in the left  arm is called $\Phi_{1x}$. 
The  flux response $\Phi_{2}$ of the variable transformer is read out by a SQUID magnetometer, magnetically coupled to the right side of the transformer with transforming ratio ${\cal R}_{2}$; the measured quantity is then the flux  $\Phi_{out}= {\cal R}_2 \Phi_2$.  
A control magnetic 
flux $\Phi_{cx}$ linked to the inner dc-SQUID modifies the device behavior. Here and in the
following, the subscript $x$ refers to externally applied signals, while $c$ refers to the control
loop.

We
introduce new coordinates given by linear combinations of the fluxes, $\varphi=\pi(\Phi_{2}-\Phi_{1}%
)/\Phi_{0}$ and $\varphi_{c}=-2\pi(\Phi_{1}+\Phi_{2})/\Phi_{0}$, where
$\Phi_{0}=h/2e$ is the flux quantum; the corresponding reduced driving fluxes
are $\varphi_{x}=-\pi\Phi_{1x}/\Phi_{0}$ and $\varphi_{cx}=2\pi\Phi_{cx}%
/\Phi_{0}$. With these coordinates, the 2D potential describing the system dynamics can be written
as follows:%

\begin{align}
U(\varphi,\varphi_{c})=\frac{\Phi_{0}^{2}}{4\pi^{2} L_{0}}\left[  \frac{1}%
{2}\left(  \varphi- \varphi_{x} \right)  ^{2} + \frac{1}{2}\gamma\left(
\varphi_{c} - \varphi_{cx}\right)  ^{2}\right. \nonumber\\
-\left.  \beta_{0} cos{\frac{\varphi_{c}}{2}}cos{\varphi}\right]
\label{eq:pot}%
\end{align} where $L_{0}=L/2+l/4$ is an effective inductance, $\beta_{0}=2\pi(2I_{0}%
)L_{0}/\Phi_{0}$ is the corresponding reduced inductance, with twice the critical current $I_{0}$
because of the two junctions in the dc-SQUID, and $\gamma=L_{0}/l$. Eq.\ref{eq:pot} has the same
form of the potential for a double SQUID~\cite{Han89}, with parameters that take into account the
gradiometric structure of the device. In the limit for an inner dc-SQUID with a vanishingly small
inductance like in our case, i.e. $\gamma>>1$, the degree of freedom related to $\varphi_{c}$ is
frozen and restrained to an equilibrium value $\varphi_{c}\simeq\varphi_{cx}$, so that the
potential
becomes a 1D curve in the remaining coordinate $\varphi$:%

\begin{equation}
U(\varphi)=E_{J}\left[  \frac{1}{2\beta}\left(  \varphi- \varphi_{x} \right)
^{2} -cos{\varphi} \right]  \label{eq:pot1D}%
\end{equation}
where $E_{J}=(2 I_{0})\Phi_{0}/(2\pi)cos(\pi\Phi_{cx}/\Phi_{0})$ and
$\beta=\beta_{0} cos(\pi\Phi_{cx}/\Phi_{0})$. This is equivalent to the potential of 
an ordinary rf-SQUID, but here the 
 critical current can be modulated by the external flux
$\Phi_{cx}$ linked to the loop of the inner dc-SQUID and the role
of reduced inductance is played by the quantity $\beta$ that is not restrained
to assume only positive values. By deriving eq.~\ref{eq:pot1D} with respect to
$\varphi$ and setting the derivative to zero, we find the relationship linking
$\varphi$ and the excitation $\varphi_{x}$ to find the extremal points:

\begin{equation}
\varphi_{x}=\varphi+[\beta_{0}cos(\pi\Phi_{cx}/\Phi_{0})]sin \varphi
\label{eq:phiphi}
\end{equation}
For $\vert\beta\vert\le1$ the relation is single-valued and the potential presents just one
minimum, while for $\vert\beta\vert> 1$ the relation is multi-valued, with 
more minima separated by potential barriers.

An  input signal centered around $\varphi_{x}=0$, with amplitude smaller than a flux quantum,
causes a monotonic and single-valued flux response $\varphi$, whose amplitude  depends on the control parameter
$\varphi_{cx}$. In a sufficiently small region this response is linear and hence the system behaves as a linear controllable transformer.  In this regime the overall transfer parameter $\cal R$, namely which part of the input magnetic flux is transmitted to the output, is given by 
the slope of the flux characteristics at the flex point $\varphi_{x}=0$.
By returning to the 
quantities $\Phi_{1x}$ (flux coupled to the left arm of the transformer), $\Phi_{2}$ (transformer response flux)
 and $\Phi_{cx}$ (flux coupled to the inner dc-SQUID), one can write:%

\begin{equation}
{\cal R}=\frac{d\Phi_{2}}{d\Phi_{1x}}\vert_{\Phi_{1x}=0}=\frac{1}{|1+\beta_{0}
cos\frac{\pi\Phi_{cx}}{\Phi_{0}}|} \label{eq:Tphi}
\end{equation}

This modulation of the overall transfer parameter $\cal R$, achieved acting on the flux $\Phi_{cx}$, is the
feature that we exploit to obtain a tunable transformer: while keeping the flux working point around zero, 
the potential is changed  in such a way to change the shape of the characteristics and operate between two points with very
different responsivity (the ``on'' and the ``off'' states). 
During operation, the system is kept 
 in the same potential well, 
 avoiding sudden dissipative jumps of the system to other minima; besides, the potential change must  be fast but still adiabatic. This last requirement represents the main limit on the operating speed of the CFC system; for our test device one can extimate this limit from  the plasma frequency $f_{p}=1/2\pi\sqrt{LC}\sim10GHz$, obtaining a value suitable for typical
superconducting quantum gates operations.

In order to test the features of the variable transformer, we built an integrated device composed
of transformer, excitation coil and non-hysteretic readout dc-SQUID, using trilayer Nb/AlOx/Nb
technology. The inner dc-SQUID~\cite{Cosmelli2001} is made by two loops in a gradiometric
configuration, with an area of $10 \mu m \times10 \mu m$, partially covered by the overlaying Nb
layer; the total inductance has been evaluated in previous measurements to be about $5 pH$. The
junctions have nominally $3 \mu m$ side and a critical current  $I_{0}\simeq 5 \mu A$,
 measured in a similar, isolated device; the corresponding reduced
inductance $\beta_{l}$, then, is much less than unity. The inner dc-SQUID is connected in parallel to two Nb
coils of inductance $L$, each consisting of  two turns  wound around a square of $100 \mu m$ side
(computed value $L=628 pH$). The transformer arms are magnetically coupled respectively to the
excitation coil and to the input coil of the readout dc-SQUID, both made of two turns, 
  nested into the rf-SQUID loops and tightly coupled by means of a ground plane; the measured transfer ratio to the readout SQUID is ${\cal R}_2= 0.20$.    For this variable transformer, the maximum
value of the reduced inductance $\beta_{0}$ is $9.5$. 

\begin{figure}
\includegraphics{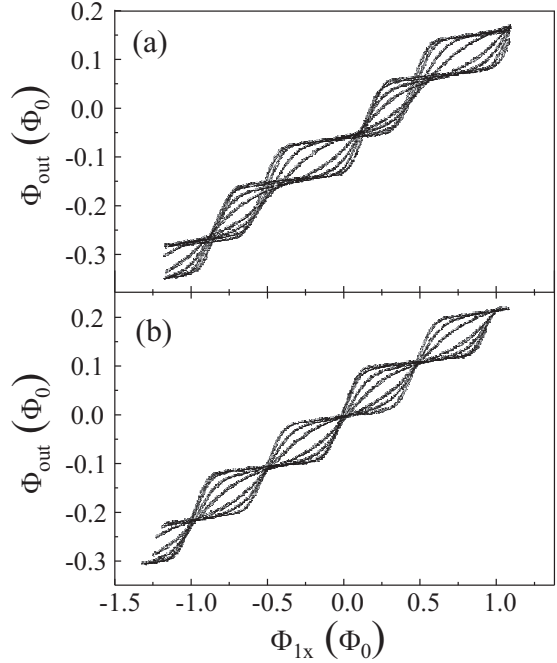}\caption{a) Experimental curves showing the flux
characteristics of the device for different values of the control magnetic
flux $\Phi_{cx}$, all in the non-hysteretic region. b) Data corrected by removing the effect of spurious couplings.}
\label{fig:caratt}
\end{figure}

In the experiment, carried out at $4.2 K$, we measured the flux response to a sweeping external flux 
for different values of the  flux $\Phi_{cx}$ applied to the inner dc-SQUID. The experimental curves
are shown in Fig.~\ref{fig:caratt}a; for clarity, curves with hysteretic behavior, which have been
observed in the device, are not displayed. The curves are described by Eq.~\ref{eq:phiphi}, except for 
a small  shift, both in the horizontal and vertical directions, caused by the unavoidable spurious coupling of the control flux $\Phi_{cx}$ to the transformer input coil and to the readout SQUID. From the measured shifts we estimated the values of the mutual inductance between the control flux coil and the transformer input   ($M_{c,1}=2.6pH$), and that between the control flux coil and the readout SQUID  ($M_{c,out}=0.29pH$).
 Fig.~\ref{fig:caratt}b shows the experimental curves after correction for the spurious coupling. 
It was verified experimentally that
the  input flux $\Phi_{1x}$  produces a negligible spurious effect both on the readout SQUID and on the inner dc-SQUID.

\begin{figure}
\includegraphics{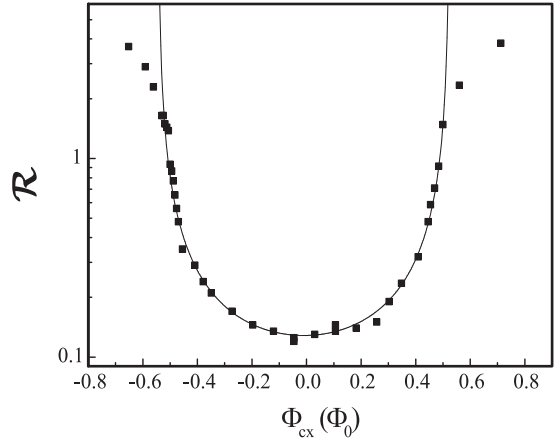}\caption{Experimental values of the slope of the flux characteristics at the working point (zero
input flux), plotted against the value of the control flux, coupled to the inner dc-SQUID. The
continuous line is the best fit with Eq.~\ref{eq:Tphi}, considering also the spurious
direct coupling between input and output. }
\label{fig:R}
\end{figure}

Fig.~\ref{fig:R} shows the measured slopes of the acquired characteristics at the working point, namely at the crossing closest to $\Phi_{1x}=0$  in Fig.~\ref{fig:caratt}a, as a function of the control flux. The fit (continuous line) is made using  Eq.~\ref{eq:Tphi}, with spurious coupling taken into account, and allows an independent estimate of $\beta_0$ that is consistent with the project value. While the agreement is very good in the bottom part of the curve, in the upper part the experimental points are lower than expected. This effect is due to the rounding of the flux characteristics because of  thermal fluctuations: the slope in the steepest points is reduced. At lower temperature this smearing effect is expected to decrease with the square root of the  temperature, until the classical-quantum crossover temperature is reached (hundreds of $mK$ for our devices) and quantum tunnelling becomes the dominant fluctuation term. The ratio between minimum and maximum slope (about $30$ for the data of Fig.~\ref{fig:R}) is a figure of merit for the performance of the variable transformer.

Let's now discuss how the variable transformer works. The ``off'' state, where the transfer ratio
is minimum, is easily identified with the state where the effective reduced inductance is maximum,
that is ${\cal R}_{min}=1/(1+\beta _{0})$; to get a small flux transfer, then, the device must
be highly hysteretic (large $\beta_{0}$). In this condition the slope of the flux characteristic is
almost horizontal and a large amount of input flux $\Phi_{1x}$ can be fed while keeping the device
in the same flux state; this curve is not shown in Fig.~\ref{fig:caratt}. By increasing the control
flux to $\Phi_{0}/2$, one gets ${\cal R}\simeq1$, and the input flux is totally transmitted to the output;
however this is not the steepest possible slope, since by further increasing $\Phi_{cx}$ one gets
$\beta$ tending to $-1$ and a diverging ${\cal R}$. Excluding the non-physical point $\beta=-1$, we can
then increase the transfer ratio ${\cal R}$ beyond unity. Two considerations must be done at this point.
First, while the ``off'' state can be obtained with a hysteretic characteristic (provided that
there are no transitions between different flux states), the ``on'' state must be obtained with
non-hysteretic characteristics and this restrains the range of usable values to $\vert
\beta\vert<1$. Second, the flux response is not necessarily linear with the input flux; the
dynamical range where linearity is ensured is depending on the parameter $\beta$: the steepest is
the slope of the flux characteristic, the smaller is the allowed flux range. With larger flux
signals, the response has a saturated amplitude and harmonics are produced. However, in certain
cases this may not be a limitation: in qubit operation, in fact, generally one has just to
distinguish between the two different flux states, the required response being just an
identification of the qubit state. In this situation, operating in the non linear range does not
affect the efficiency of the measurement and extends the usable range of the input signals. As a
matter of fact, choosing the working points for ``off'' and ``on'' states is best achieved
experimentally, according to the specific requirements of the experiment and to the signal
characteristics that must be preserved.

To test the performance of the transformer, we sent a sinusoidal signal to the
left side of the transformer and measured the output from the readout dc-SQUID
while modulating the inner dc-SQUID with a square wave between the points of
maximum and minimum transfer ratio, chosen experimentally.

\begin{figure}
\includegraphics{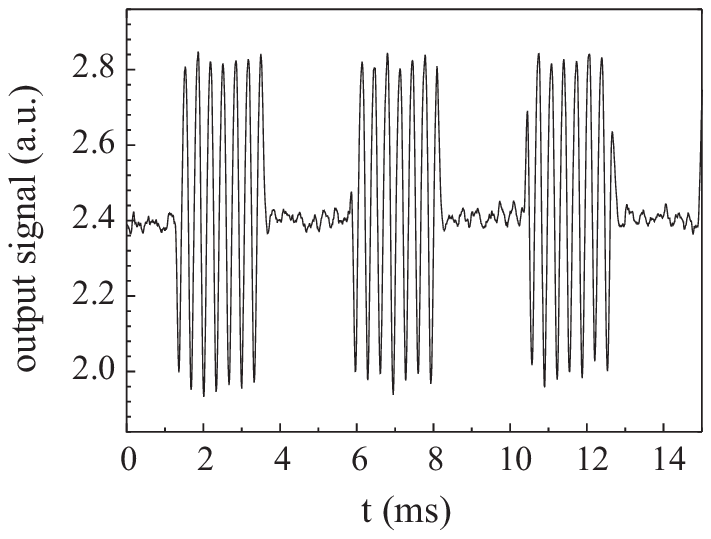}\caption{Modulation of a sinusoidal signal fed on one
side of the tunable transformer. The ratio of the transmitted amplitude
between on and off states is 30.}
\label{fig:sin}
\end{figure}
The resulting modulation is shown in Fig.~\ref{fig:sin}.The measured efficiency, the maximum ratio between the ``on'' and the ``off'' transmission, is
about $30$, in agreement with  the data of Fig.~\ref{fig:sin}. For this situation, the measured range in which the response is linear corresponds a peak-to-peak input flux of $0.1 \Phi_{0}$. If the linear range is exceeded, the efficiency in the flux modulation decreases but
operation is still possible. As an example, for an input flux of $0.3 \Phi_{0}$ peak-to-peak the
measured ratio between the ``on'' and the ``off'' transmission is about $14$.

\begin{figure}
\includegraphics{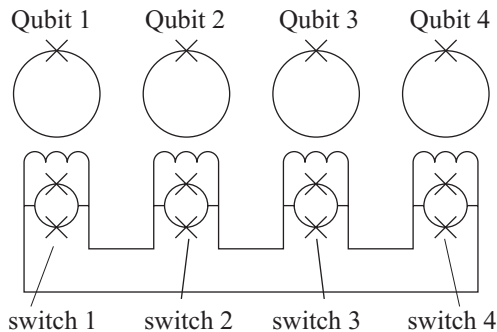} 
\caption{Example of a possible scheme for the controlled
coupling of more flux qubits.}
\label{fig:bus}
\end{figure}

CFC systems  can be integrated together with the SQUID flux qubits, since they are based on the same
technology, and can be used to control the couplings between them. They can also be also used to
realize a bus for the controlled coupling of more qubits, for example by using the scheme shown in
Fig.~\ref{fig:bus}; in this example a pair of qubits can be coupled by switching ``on'' the
relative ``switches'', and by maintaining all the others in the ``off'' state (this scheme is
similar to that proposed in ~\cite{chiarello}). Since the CFC remains always in the superconducting
state without jumps to a dissipative state, the only contribution to the overall decoherence is due
to the device intrinsic dissipation and to the enviromental noise pick-up. This means that the
total contribution to decoherence should be of the same order of the qubit contribution, since they
are very similar for technology, dimensions, components and structure.

In conclusion, we have realized and tested a microfabricated SQUID based
controllable flux coupling, useful in any application in which it is necessary
to modify the magnetic coupling between different devices, and in particular
suitable for quantum computing applications with flux qubits.

\ \ 
This work has been supported by INFN under the project SQC.

\end{document}